\documentclass[iop,apj]{emulateapj}
\submitted{ApJ, in press}

\newcommand{\ergs}{\,erg\,s$^{-1}$}
\newcommand{\ergcm}{\,erg\,cm$^{-2}$\,s$^{-1}$}

\newcommand{\pdot}{\mbox{$\dot{P}$}}
\newcommand{\edot}{\mbox{$\dot{E}$}}
\newcommand{\cxo}{\emph{Chandra}}
\newcommand{\xmm}{\emph{XMM}} 
\newcommand{\fer}{\emph{Fermi}} 
\newcommand{\nh}{$N_{\rm H}$}
\newcommand{\cms}{\,cm$^{-2}$}
\newcommand{\psr}{PSR B1937+21}
\newcommand{\blc}{$B_{\rm lc}$}

\begin{document}
\title{HIGH-ENERGY EMISSION OF THE FIRST MILLISECOND PULSAR}

\begin{abstract}
We report on X-ray and gamma-ray observations of the millisecond pulsar
(MSP) B1937+21 taken with the \emph{Chandra X-ray Observatory},
\emph{XMM-Newton}, and the \emph{Fermi} Large Area Telescope. The pulsar X-ray
emission shows a purely non-thermal spectrum with a hard photon index of
$0.9\pm0.1$, and is nearly 100\% pulsed. We found no evidence of varying pulse
profile with energy as previously claimed. We also analyzed 5.5\,yr of \fer\
survey data and obtained much improved constraints on the pulsar's timing and
spectral properties in gamma-rays. The pulsed spectrum is adequately fitted by
a simple power-law with a photon index of $2.38\pm0.07$. Both the gamma-ray
and X-ray pulse profiles show similar two-peak structure and generally align
with the radio peaks. We found that the aligned profiles and the hard spectrum
in X-rays seem to be common properties among MSPs with high magnetic fields at
the light cylinder. We discuss a possible physical scenario that could give
rise to these features.
\end{abstract}

\author{C.-Y. Ng\altaffilmark{1}, J. Takata\altaffilmark{1}, G. C. K.
Leung\altaffilmark{1}, K. S. Cheng\altaffilmark{1}, and P.
Philippopoulos\altaffilmark{2}}
\altaffiltext{1}{Department of Physics, The University of Hong Kong, Pokfulam
Road, Hong Kong, China}
\altaffiltext{1}{Department of Physics, McGill University, Montreal, QC H3A
2T8, Canada}

\email{ncy@bohr.physics.hku.hk}
\shorttitle{High-Energy Emission of \psr}
\shortauthors{Ng et al.}


\section{INTRODUCTION}
Millisecond pulsars (MSPs) are fast-spinning neutron stars with rotation
periods ($P$) from a few to tens of milliseconds. They are believed to be old
pulsars spun up through accretion process from a companion star.
MSPs generally show small spin-down rates (\pdot) that imply low surface
magnetic fields ($B_s\propto\sqrt{P\dot P}$) of the order of $10^8$\,G, much
lower than those of young pulsars. Nonetheless, MSPs exhibit broadband
emission as young pulsars do, and are detected across the electromagnetic
spectrum from radio to X-ray to gamma-ray bands. Recently there has been
significant progress in the high-energy studies of MSPs. The Large Area
Telescope (LAT) onboard the \emph{Fermi Gamma-ray Space Telescope} has
detected over 60 MSPs in gamma-rays since 2008 \citep[see the LAT second
pulsar catalog;][]{aaa+13}. In addition, the \emph{Chandra X-ray Observatory}
and the \emph{XMM-Newton} mission have provided sensitive measurements of the
X-ray properties of over a dozen MSPs \citep[see ][and references
therein]{hnk13}. These have significantly expanded the sample and allowed
detailed studies of the population.

In this study we focus on the X-ray and gamma-ray emission properties of
\object{\psr}, which is a representative of an emerging class of MSPs that
shows aligned pulse profiles in different energy bands \citep{gjv+12}. This
could possibly indicate a different emission mechanism than that of typical
pulsars. \psr\ (also known as PSR J1939+2134) is the first MSP discovered
\citep{bkh+82} and it remains the second fastest-spinning pulsar known, with
$P=1.56$\,ms and $\dot P=1.05\times10^{-19}$. (Corrections of \pdot\ due to
proper motion and differential Galactic rotation are negligible). These timing
parameters suggest $B_s=4.1\times10^8$\,G and spin-down power $\dot E=4\pi^2
I\dot P/P^3=1.1\times10^{36}$\ergs, where $I=10^{45}$\,g\,cm$^2$ is neutron
star moment of inertia. This is the second largest \edot\ among MSPs, only
after PSR B1821$-$24.

At high energies, \psr\ was first detected in X-rays with \emph{ASCA}
\citep{tst+01}. The emission has a non-thermal spectrum and exhibits strong
pulsations. The pulse profile shows a narrow peak structure with a pulsed
fraction (PF) of 44\%. The study also suggests a hint of a second peak in the
profile. Subsequent observations with \emph{RXTE} and \emph{BeppoSAX}
confirmed the second peak \citep{chk+03,ncl+04}, and it was claimed that the
relative strength of the two peaks and the PF could vary with energy, from
PF$=85\%\pm5$\% in 1.3--10\,keV to PF$=54\%\pm7$\% in 4--10\,keV \citep{ncl+04}.
\citet{zav07} reported on a \cxo\ observation of the source and found
an X-ray photon index $\simeq1.2$. Using 1.5\,yr of \fer\ survey data,
\citet{gjv+12} detected gamma-ray pulsations from \psr. The pulse
profile well aligns with the radio profile, indicating that both emission
could originate from the same region in the outer magnetosphere. Spectral
analysis using the \fer\ data suggests that the phase-averaged pulsar spectrum
could be fitted with an exponentially cutoff power-law model \citep{gjv+12}.
However, note that the spectral parameters of this source are not listed in
the LAT second pulsar catalog that uses three years of data, due to low
detection significance with a test-statistic (TS) value of 10 only
\citep{aaa+13}.

\begin{deluxetable*}{lcccccc}
\tablecaption{\cxo\ and \xmm\ Observations of \psr \label{table:obs}}
\tablewidth{5.3in}
\tablehead{
\colhead{Telescope} & \colhead{Observation} & \colhead{ObsID} &
\colhead{Instrument} & \colhead{Mode} & \colhead{Time Reso-} &
\colhead{Net Expo-}\\
& \colhead{Date} & & & &
\colhead{lution\ (s)} &
\colhead{sure (ks)} }
\startdata
\cxo & 2005 Jun 28 & 5516 & ACIS-S  &  Faint & 3.2 & 49.5 \\
\xmm & 2010 Mar 29 & 0605370101 & MOS1 & Full Frame & 2.6 & 40.4 \\
& & & MOS2 & Full Frame & 2.6 & 47.5 \\
& & & PN & Timing & $3\times10^{-5}$ & 40.4
\enddata
\end{deluxetable*}

We present a new study of the high-energy emission of \psr\ using archival
X-ray data made with \cxo\ and \xmm\ and 5.5\,yr of \fer\ survey data. The
observations and data reduction are described in Section~\ref{s2} and the
analysis and results are presented in Section~\ref{s3}. In Section~\ref{s4},
We compare the results with other MSPs and discuss possible physical emission
mechanisms. Our findings are summarized in Section~\ref{s5}.

\section{OBSERVATIONS AND DATA REDUCTION} \label{s2}
We reprocessed the archival \cxo\ and \xmm\ data. The former, which have been
used in the previous study \citep{zav07}, were taken on 2005 Jun 28 using the
ACIS-S detector in the imaging mode and have a time resolution of 3.2\,s. The
\xmm\ observation was taken on 2010 March 29 with the MOS1 and MOS2 cameras in
the full frame mode that has a frame time of 2.6\,s and the PN camera in the
timing mode that has a high time resolution of 0.03\,ms. We performed the
\cxo\ and \xmm\ data reduction using CIAO 4.5 and SAS 12.0, respectively. In
the \xmm\ analysis, only \texttt{PATTERN} $\leq 12$ events from MOS and
\texttt{PATTERN} $\leq 4$ events from PN were used to ensure good data
quality. After removing periods of high background, we obtained net exposures
of 49.5\,ks, 40.4\,ks, 47.5\,ks, and 40.4\,ks from \cxo, MOS1, MOS2, and PN,
respectively. The observation parameters are listed in Table~\ref{table:obs}.

For the gamma-ray analysis, we selected \fer\ LAT Pass 7 reprocessed data
\citep[P7REP;][]{bcw+13} taken between 2008 August 4 and 2014 January 17.
Class 2 events in the P7REP\_SOURCE\_V15 instrument response function (IRFs)
were used throughout this paper, and the data reduction was carried out with
the \fer\ Science Tools v9r32p5. We restricted the analysis on data with
zenith angles less than 100\arcdeg, with spacecraft rocking angle less than
52\arcdeg, and in the 0.1--100\,GeV energy range.

\section{ANALYSIS AND RESULTS}\label{s3}
\begin{figure}[!ht]
\epsscale{1.2}
\plotone{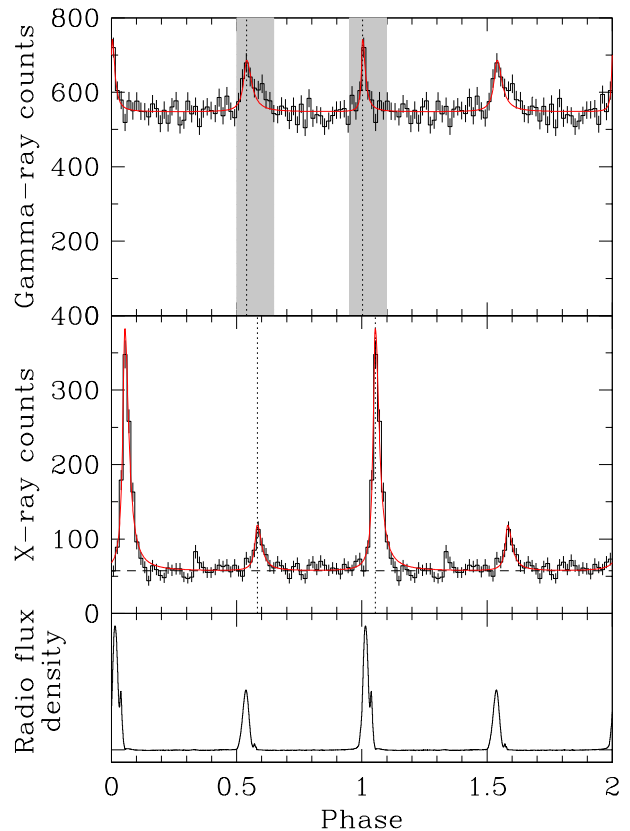}
\caption{Pulse profiles of \psr\ in gamma-rays (0.1--100\,GeV; upper panel),
X-rays (0.5--7\,keV; middle panel), and radio (1.4\,GHz; lower panel). The
latter is from \citet{gjv+12}. The best-fit models with two asymmetric
Lorentzians are shown by the red solid lines. The vertical dotted lines
indicate the best-fit peak positions and the horizontal dashed line shows the
X-ray background level. The shaded regions indicate the on-pulse phase
intervals used for \fer\ spectral analysis. (A color version of this figure is
available in the online journal.)
\label{fig:profile}}
\end{figure}
\subsection{X-Ray Analysis}
\psr\ is clearly detected in both the \cxo\ and the \xmm\ observations, and
the radial profiles of the images are consistent with a point source, same as
what \citet{zav07} found. Using a 3\arcsec\ radius aperture centered on the
pulsar, we found $610\pm25$ background-subtracted \cxo\ counts from the in the
0.5--7\,keV energy range. Similarly, we extracted $370\pm20$ and $430\pm22$
net counts from MOS1 and MOS2, respectively, using a 20\arcsec-radius
aperture. The PN data have higher background since they were taken in the
timing mode. We obtained $1230\pm70$ source counts from a 33\arcsec-wide
rectangular region in 0.5--7\,keV.

\subsubsection{Timing}
We performed timing analysis with only the PN data, as they have a high time
resolution. The photon arrival times were first corrected to the Solar system
barycenter, then folded using the TEMPO2 photons
plug-in\footnote{\url{http://www.physics.mcgill.ca/$\sim$aarchiba/photons\_plug.html}}
with the ephemeris from radio timing\footnote{The timing parameter
file is obtained from the LAT second pulsar catalog
\url{http://fermi.gsfc.nasa.gov/ssc/data/access/lat/2nd\_PSR\_catalog/}.} \citep{sgc+08}.
The resulting X-ray profile in the 0.5--7\,keV energy range is shown in
Figure~\ref{fig:profile}. We have tried other energy bands between 0.5 and
10\,keV, and found no energy dependence. The profile exhibits very sharp main
pulse and interpulse components $\sim$180\arcdeg\ apart. In addition, there is
a hint of a third peak in between, albeit it is not statistically significant.
A direct comparison with the background level obtained from a nearby region
indicates that the emission is nearly 100\% pulsed. We also plotted the
1.4\,GHz radio profile in the figure for comparison. The X-ray
pulse and interpulse slightly lag the radio ones.

To quantitatively measure the peak position and width, we followed
\citet{aaa+13} to fit the pulse profile using an unbinned maximum likelihood
method. We employed a simple model with two asymmetric Lorentzians. The
functional form is given by
\begin{equation}
g(x)= \frac{2}{\pi}\frac{A_1}{(\sigma_1^-+\sigma_1^+)(1+z_1^2)}
+\frac{2}{\pi}\frac{A_2}{(\sigma_2^-+\sigma_2^+)(1+z_2^2)}~,
\end{equation}
where
\begin{equation}
z_i = \left\{
\begin{array}{lr}
(x-\Phi_i) / \sigma_i^- & \mbox{if } x\leq \Phi_i \\
(x-\Phi_i) / \sigma_i^+ & \mbox{if } x > \Phi_i 
\end{array}
\right .~,
\end{equation}
$A_i$ are the amplitudes, $\Phi_i$ are the peak positions, and $\sigma_i^\pm$
are the width parameters such that the full widths at half-maximum (FWHMs) of
the two peaks are given by $\sigma_1^-+\sigma_1^+$ and
$\sigma_2^-+\sigma_2^+$. We found that the main pulse and interpulse peak at
phase $0.051\pm0.001$ and $0.583^{+0.006}_{-0.005}$,
respectively.\footnote{Phase 0 is defined by the maximum of the first Fourier
harmonic of the radio profile in the time domain \citep{gjv+12}.} The
best-fit results with statistical uncertainties at the 1$\sigma$ level are
listed in Table~\ref{table:results} and the model profile is plotted in
Figure~\ref{fig:profile}. The uncertainties are estimated using Monte Carlo
simulations. We generated 1000 random realizations of the best-fit model and
fitted each one to obtain distributions of the best-fit parameters. The
confidence intervals are determined from the most compact regions that contain
68\% of the sample. Note that we did not account for any systematic
uncertainties, such as the absolute timing accuracy of the instrument.
Finally, we searched for bursts that may correspond to the giant pulses and
any flux variabilities at longer timescale, but found negative result.

\begin{deluxetable}{lccc}
\tablecaption{Best-fit Timing and Spectral Parameters of \psr
\label{table:results}}
\tablewidth{0pt}
\tablehead{
\colhead{Parameter} & \colhead{X-ray} & \colhead{Gamma-ray}}
\startdata
\cutinhead{Timing}
\sidehead{First peak}
Position, $\Phi_1$ & $0.054\pm0.001$ & $0.003^{+0.001}_{-0.003}$ \\
Width parameter, $\sigma_1^-$ & $0.010^{+0.001}_{-0.002}$
 & $0.006^{+0.001}_{-0.004}$ \\
Width parameter, $\sigma_1^+$ & $0.019^{+0.001}_{-0.003}$
 & $0.012^{+0.001}_{-0.005}$ \\
FWHM$_1=\sigma_1^-+\sigma_1^+$ & $0.029^{+0.001}_{-0.003}$
 & $0.019^{+0.002}_{-0.006}$ \\
Radio lag, $\delta_1$ & $0.068\pm0.001$ & $-0.011^{+0.001}_{-0.004}$ \\
\sidehead{Second peak}
Position, $\Phi_2$ & $0.583\pm0.005$ & $0.540^{+0.006}_{-0.005}$ \\
Width parameter, $\sigma_2^-$ & $0.012^{+0.003}_{-0.007}$ &
$0.015^{+0.001}_{-0.007}$ \\
Width parameter, $\sigma_2^+$ & $0.020^{+0.001}_{-0.009}$ &
$0.023^{+0.001}_{-0.010}$ \\
FWHM$_2=\sigma_2^-+\sigma_2^+$ & $0.032^{+0.003}_{-0.012}$ &
$0.018^{+0.002}_{-0.012}$ \\
Radio lag, $\delta_2$ & $0.032_{-0.005}^{+0.006}$ & $0.003^{+0.006}_{-0.005}$ \\
\cutinhead{Spectroscopy}
Column density, \nh\ ($10^{22}$\cms) & $1.2\pm0.2$ & \nodata \\
Photon index, $\Gamma$ & $0.9\pm0.1$ & $2.38\pm0.07$ \\
Statistic & $\chi^2_\nu=0.86$ & TS-value=112\\
Unabsorbed energy flux\tablenotemark{a} ($10^{-12}$ &
$0.23_{-0.03}^{+0.04}$ & $16\pm2$\\
 \ergcm) && \\
Luminosity\tablenotemark{a} ($10^{33}$\ergs) & 0.7 & 49 \\
Efficiency\tablenotemark{a}, $\eta$ & $6\times10^{-4}$ & 0.044
\enddata
\tablenotetext{a}{Phase-averaged values assuming no off-pulse emission. Also,
no beaming correction has been applied. The X-ray and gamma-ray energy
ranges are 0.5--7\,keV and 0.1--100\,GeV, respectively.}
\tablecomments{All uncertainties are statistical errors at the 1$\sigma$
confidence level. The timing parameters are in phase units between 0 and 1.}
\end{deluxetable}

\begin{figure}[ht]
\epsscale{1.3}
\plotone{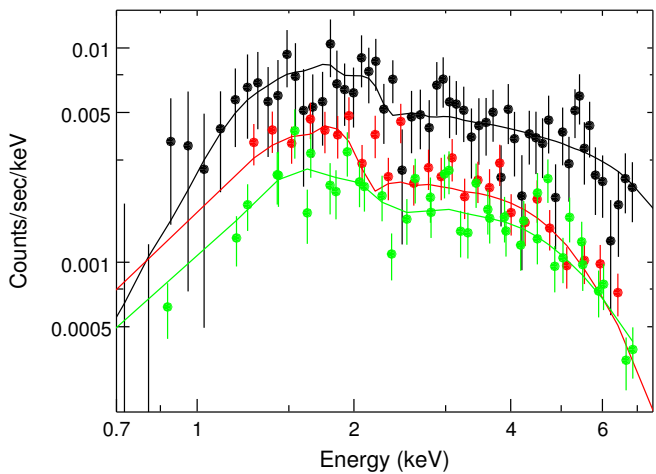}
\caption{X-ray spectrum of \psr. The \cxo, PN, and MOS1+MOS2 data are
shown in red, black, and green, respectively. The solid lines indicate
the best-fit absorbed power-law model.
(A color version of this figure is available in the online journal.)
\label{fig:spec}}
\end{figure}

\subsubsection{Spectroscopy}
We extracted the pulsar spectrum from the regions stated above. Backgrounds
were obtained from nearby regions on the same chip. As the pulsar is 100\%
pulsed, we extracted the PN spectrum from the on-pulse phase intervals only,
between phase 0--0.15 and 0.52--0.65, in order to boost the signal-to-noise
ratio. Note that there is a faint source CXO J193939.3+213506 located
13\arcsec\ northeast of the pulsar, which is not resolved by \xmm. However,
there should be negligible contamination to the pulsar spectrum, since its
\cxo\ count rate in 0.5--7\,keV is only 2\% of that of the pulsar. This is
also supported by the $\sim$100\% PF of the PN counts.

The spectral fitting was performed in the Sherpa environment. We grouped the
ACIS and MOS spectra to at least 20 counts per bin, and the PN spectrum to at
least 30 counts per bin. All four spectra (ACIS, MOS1, MOS2, and PN) were
fitted jointly in the 0.5--7\,keV energy range. We employed a simple absorbed
power-law model with abundances and the absorption cross sections given by
\citet{wam00}. This provides a good fit with a reduced $\chi^2$ value of 0.86
over 122 degrees of freedom. The best-fit parameters are listed in
Table~\ref{table:results} and the model is shown in Figure~\ref{fig:spec}. The
fit gives a column density of $N_{\rm H}=(1.2\pm 0.2)\times10^{22}$\cms\ and a
photon index of $\Gamma=0.9\pm0.1$ (all uncertainties are at the $1\sigma$
confidence level). The absorbed and unabsorbed fluxes in 0.5--7\,keV are
$(1.8\pm0.3)\times10^{-13}$\ergcm\ and $2.3_{-0.3}^{+0.4}\times
10^{-13}$\ergcm, respectively. For the source distance of 5\,kpc
\citep{vwc+12}, this converts to X-ray luminosity of $L_X=6.8\times10
^{32}$\ergs, implying an efficiency of $L_X/\dot E=6\times10^{-4}$.
To check the cross-calibration, we fit the \cxo\ and \xmm\ data separately
and obtained $N_{\rm H}=1.3_{-0.4}^{+0.5}\times10^{22}$\cms\
with $\Gamma_X=0.8\pm0.1$ from the former
and $N_{\rm H}=(1.1\pm 0.2)\times10^{22}$\cms\
with $\Gamma_X=1.0\pm0.2$ from the latter. These values are
fully consistent.

We have also tried a blackbody model, but the fit is slightly worse
($\chi^2_\nu=0.95$) and the temperature seems too high ($kT=1.5$\,keV) to be
physical. Moreover, adding a blackbody component to the power-law model shows
no significant improvement to the fit. Nonetheless, we can still post a limit
on the thermal emission by adjusting the blackbody parameters until the
$\chi^2$ value exceeded a certain level. We obtained a temperature limit of
$kT<0.13$\,keV at the 99\% confidence level, for any thermal emission from the
polar cap region with a radius $r_{\rm pc}=R_\ast\sqrt{2\pi R_\ast /
cP}=3.7$\,km, where $R_\ast=10$\,km is the neutron star radius.

\subsection{Gamma-Ray Analysis}
\subsubsection{Timing}
Using a 1\arcdeg\ radius region of interest (ROI) centered on the pulsar
position, $3.7\times10^4$ events in the 0.1--100\,GeV energy range were
extracted from the \fer\ observation. We applied a barycentric correction to
the photon arrival times and folded them using the TEMPO2 fermi plug-in, with
the same radio ephemeris as for the X-ray analysis. Pulsations are clearly
detected and the $H$-test \citep{db10} gives a statistic of 78, which
corresponds to 7.6$\sigma$ detection. The pulse profile is shown in
Figure~\ref{fig:profile}. We have also tried the 1--100\,GeV energy range and
the profile looks very similar, albeit with a lower $H$-statistic of 19 only.

The gamma-ray profile in the figure shows sharp main pulse and interpulse
that resemble the X-ray ones, but the gamma-ray pulses better align with the
radio peaks. We fitted the gamma-ray profile with the same algorithm as
described in the X-ray analysis above and the best-fit parameters and
statistical uncertainties are listed in Table~\ref{table:results}. 
Finally, we checked the long term gamma-ray flux evolution of the
source, and found no significant variability.

\subsubsection{Spectroscopy}
We performed a binned likelihood analysis using a $20\arcdeg\times20\arcdeg$
square ROI. Our model includes all sources in the Second Fermi-LAT Source
Catalog \citep[2FGL;][]{naa+12} within 20\arcdeg\ from the pulsar, the
Galactic diffuse emission (gll\_iem\_v05.fits), and the extragalactic
isotropic emission (iso\_source\_v05.txt). The source spectral models are all
adopted from 2FGL. For sources located more than 8\arcdeg\ from the pulsar,
their spectral parameters are held fixed during the fit. We have tried a
power-law model and a power-law with exponential cutoff model (PLEC) for the
pulsar spectrum. The latter has a functional form of
\begin{equation}
\frac{dN}{dE}\propto E^{-\Gamma}\exp\left(-\frac{E}{E_c}\right)~,
\end{equation}
where $N$ is the photon flux and $E_c$ is the cutoff energy. We found that
both models give low TS values that correspond to a detection significance
below 3$\sigma$ level.
\begin{figure}[th]
\epsscale{1.3}
\plotone{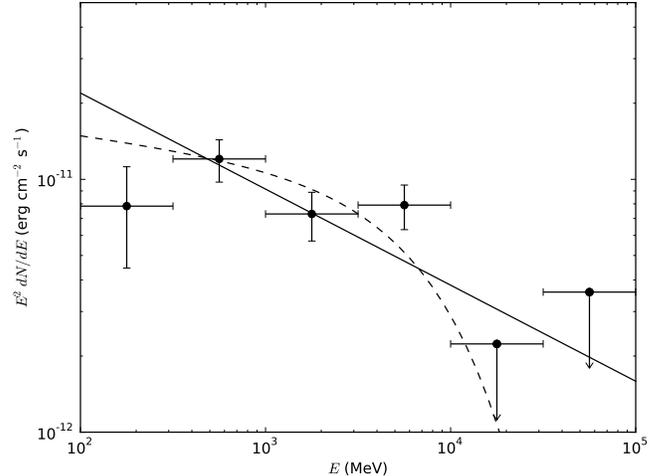}
\caption{Pulsed \fer\ gamma-ray spectrum of \psr\ obtained from binned maximum
likelihood analysis. The best-fit power-law and power-law with exponential
cutoff models are indicated by the solid and dashed lines, respectively.
Statistical uncertainties at the 1$\sigma$ confidence level are shown here. The
upper limits are at the 2$\sigma$ confidence level.
\label{fig:fermi}}
\end{figure}

\begin{deluxetable*}{lccccccccc}
\tablecaption{High-\edot\ MSPs and Their X-ray Properties.\label{table:highedot}}
\tablewidth{5in}
\tabletypesize{\footnotesize}
\tablehead{
\colhead{Name} & \colhead{$P$}&\colhead{\edot\ $(10^{35}$}&\colhead{$B_s$} &
\colhead{\blc} & \colhead{Dist.} & \colhead{$\Gamma_X$} &
\colhead{$L_X$ $(10^{32}$} & \colhead{PF} & \colhead{Ref.}\\
& \colhead{(ms)} & \colhead{\ergs)} & \colhead{($10^8$\,G)} &
\colhead{($10^5$\,G)} & \colhead{(kpc)} & & \colhead{\ergs)} & \colhead{(\%)}&}
\startdata
B1821$-$24 &3.1&22&23&7.3&5.1&$1.23\pm0.03$&14&$82.5\pm4$&1\\
B1937+21 &1.6&11 &4.1 &10& 5.0 & $0.9\pm0.1$&6.8&$\sim100$ & This work \\
B1820$-$30A&5.4&8.3\tablenotemark{a}&43\tablenotemark{a}&2.5\tablenotemark{a}&
 7.9&\nodata&\nodata&\nodata&\nodata\\
J1701$-$3006F&2.3&7.3\tablenotemark{a}&7.2\tablenotemark{a}&
 5.5\tablenotemark{a}&6.9&\nodata&\nodata&\nodata&\nodata\\
J1701$-$3006E&3.2&3.6\tablenotemark{a}&10\tablenotemark{a}&
 2.7\tablenotemark{a}&6.9&\nodata&\nodata&\nodata&\nodata\\
J0218+4232 &2.3&2.4&4.3 &3.1& 2.7 & $1.10\pm0.06$ &3.3&$64\pm6$&2, 3 \\
B1957+20&1.6&1.1&1.4 &3.0& 2.5 & $\sim2$\tablenotemark{b}&
 $\sim0.5$\tablenotemark{b}&\nodata& 4\\
J1750$-$3703D &5.1&1.4\tablenotemark{a}&16\tablenotemark{a}&
 1.1\tablenotemark{a}& 12 & $<50$ &\nodata &\nodata& 5\\
B0021$-$72F&2.6&1.4&4.1&2.1&4.0&\nodata&0.055\tablenotemark{c}& \nodata& 6\\
J1740$-$5340A&3.7&1.4\tablenotemark{a}&7.9\tablenotemark{a}&1.5\tablenotemark{a}&
 3.4&$1.73\pm0.08$\tablenotemark{b} & 0.22\tablenotemark{b}&\nodata& 7\\
J1701$-$3006D&3.4&1.2\tablenotemark{a}&6.6\tablenotemark{a}&1.5\tablenotemark{a}&
 6.9&\nodata&\nodata&\nodata&\nodata
\enddata
\tablenotetext{a}{Observed value --- there is no correction to \pdot\ for
the pulsar proper motion or differential Galactic rotation.}
\tablenotetext{b}{The X-ray emission could be dominated by intra-binary
shock (see the references).}
\tablenotetext{c}{The X-ray spectrum is best fitted with a thermal model.}
\tablerefs{(1) \citealt{bvs+11}; (2) \citealt{wob04}; (3) \citealt{khv+02};
(4) \citealt{hkt+12}; (5) \citealt{bhp10}; (6) \citealt{bgh+06};
(7) \citealt{bvh+10}.}
\end{deluxetable*}

To boost the signal, we restricted the analysis on the on-pulse data only, in
phase 0.5--0.65 and 0.95--1.1. These intervals are indicated in
Figure~\ref{fig:profile}. We obtained a much higher TS value of 112, i.e.,
above 10$\sigma$ significance, using the power-law model. The fit gives a
photon index of $\Gamma_\gamma=2.38\pm0.07$. The best-fit parameters and the
phase-averaged energy flux are listed in Table~\ref{table:results}. We have
also tried the PLEC model and found a sightly better fit (TS = 116) with
$\Gamma_\gamma=2.1\pm0.2$ and $E_c=8\pm4$\,GeV. However, the likelihood ratio
test gives $-2\Delta\log(\mbox{likelihood})=\Delta \mbox{TS} <9$. We therefore
follow the convention in \citet{aaa+13} to conclude that the improvement is
not significantly preferred. The best-fit power-law and PLEC models are
plotted in Figure~\ref{fig:fermi}. 

\section{DISCUSSION} \label{s4}
\subsection{Comparison with Previous Studies}
The \cxo\ and \xmm\ observations offer better angular resolution and
sensitivity than previous X-ray studies with \emph{ASCA}, \emph{RXTE}, and
\emph{BeppoSAX}, providing substantial improvements on the spectral and
timing measurements. While $\Gamma_X$ we obtained is consistent with the
reported values, we found a lower $N_\mathrm{H}$ of $1.2\times10^{22}$\cms\
compared to $\sim2\times10^{22}$\cms\ given by \citet{tst+01}, \citet{chk+03},
and \citet{ncl+04}. Our result is closer to the total Galactic H{\sc i} column
density of $1.1\times 10^{22}$\cms\ in the direction measured with radio
observations \citep{kbh+05}. We however note that it is not unusual for sources
near the Galactic plane to have an X-ray absorption column density larger than
the total H{\sc i} column density, because X-rays are mostly absorbed by
molecular clouds instead of neutral hydrogen atoms \citep[see][]{hnk13}. The
small \nh\ value results in slight smaller unabsorbed flux of $3.0\times
10^{-13}$\ergcm\ in 2--10\,keV, as compared to $\sim4\times10^{-13}$\ergcm\
in \citet{tst+01} and \citet{ncl+04}.

For the timing analysis, the \xmm\ PN observation allows us to isolate the
pulsar flux from the background emission, revealing for the first time the
near 100\% PF of the X-ray emission. This is much higher than $\sim$60\% found
with \emph{ASCA} \citep{tst+01}. Our results show no evidence of
energy-dependent pulse profile in the soft X-ray band, thus, rejecting the
claim of a lower ($54\%\pm7$\%) PF below 4\,keV \citep{ncl+04}. The X-ray peak
positions we obtained are mostly consistent with the published values. While a
direct comparison with the \emph{RXTE} results may seem to indicate slight
offsets ($\Delta \phi_1=0.0028\pm 0.001$, $\Delta \phi_2=0.0078\pm 0.005$), we
note that these are smaller than the 40\,$\mu$s absolute timing accuracy of
\xmm\ \citep{mkc+12}, which corresponds to 0.03 in phase units.

\psr\ is one of the few pulsars with radio giant pulses detected.
\citet{chk+03} noticed phase alignment between the radio giant and X-ray
pulses, suggesting a possible correlation. Although there is no
contemporaneous radio observation accompanied with the \xmm\ timing data, a
large number of giant pulses are expected during the 11\,hr long X-ray
exposure, given their observed rate of a few hundreds to a few thousands per
hour \citep[e.g.,][]{spb+04, zps+13}. We found no clustering of X-ray photon
arrival times, same as what \citet{chk+03} concluded. Similar negative results
have been reported for other pulsars that show giant pulses
\citep[e.g.,][]{jrm+04,bmk+12}.

Thanks to more \fer\ data and the improved IRF \citep[see][]{bcw+13}, our
gamma-ray timing results are fully consistent with, but better constrained
than, those from \citet{gjv+12}. For the spectral analysis, we found a
slightly harder photon index $\Gamma_\gamma=2.38\pm0.07$ compared to
$3.02\pm0.18$ \citep{gjv+12}. On the other hand, our PLEC fit gives a softer
$\Gamma_\gamma$ ($2.1\pm0.2$ versus $1.43\pm0.87$), although the two are
formally compatible given the large uncertainties of the latter. We also
obtained a higher cutoff energy at $E_c=8\pm4$\,GeV as compared to
$E_c=1.2\pm0.7$\,GeV. However, we note that $E_c$ is not very well determined
in both cases.

\subsection{High-\edot, High-\blc\ MSPs}
The non-thermal spectrum and large PF of the X-ray emission of \psr\ indicate
a magnetospheric origin. These characteristics seem common among high-\edot\
MSPs \citep[see][]{zav07}. We list in Table~\ref{table:highedot} all known
MSPs with $\dot E>10^{35}$\ergs. Three of them, PSRs B1821$-$24, B1937+21, and
J0218+4232, have confirmed X-ray pulsations.\footnote{\citet{gjv+12} reported
evidence of X-ray pulsations from PSR B1957+20, but the overall X-ray emission
could be dominated by intra-binary shock emission \citep{hkt+12}.} They all
exhibit hard power-law spectrum in X-rays with $\Gamma_X\sim 1$, large PF of
64\%--100\%, general alignment between the X-ray and radio pulses
\citep{khv+02, jgk+13}, and giant radio pulses \citep{rj01,wcs84,kbm+06}. It
has been suggested that these properties could be attributed to a strong
magnetic fields at the light cylinder. Its strength, \blc, is related to the
surface dipole field $B_s$ by $B_{\rm lc}=B_s(R_\ast/R_{\rm lc})^3$, where
$R_{\rm lc}=Pc/2\pi$ is the light cylinder radius. All MSPs listed in the
table have strong \blc\ above $10^5$\,G. In particular, \psr\ has the highest
\blc\ of $1.0\times 10^6$\,G, even higher than that of the Crab pulsar.

\begin{deluxetable*}{lcccccc}
\tablecaption{Properties of \fer-LAT detected MSPs
\label{table:fermimsp}}
\tablewidth{5in}
\tablehead{
\colhead{Name} & \colhead{$P$}&\colhead{\edot\ $(10^{34}$}&
 \colhead{$B_s$} & \colhead{\blc} & \colhead{X-ray} &
 \colhead{Radio/Gamma-} \\
& \colhead{(ms)} & \colhead{\ergs)} & \colhead{($10^8$\,G)} &
 \colhead{($10^5$\,G)} & \colhead{Spectrum\tablenotemark{a}} &
 \colhead{Ray Class\tablenotemark{b}} }
\startdata
B1937+21&1.6&110&4.1&9.9&PL &II\\
B1821$-$24&3.1&220&23&7.3&PL&II\\
J0218+4232&2.3&24&4.3&3.2&PL&II\\
J1747$-$4036&1.7&12\tablenotemark{c}&1.5\tablenotemark{c}&
 3.1\tablenotemark{c}&\nodata&I/II\\
B1957+20&1.6&7.6&1.2&2.5&\nodata&II\\
B1820$-$30A&5.4&83\tablenotemark{c}&43\tablenotemark{c}&
 2.5\tablenotemark{c}&\nodata&II\\
J1902$-$5105&1.7&6.7\tablenotemark{c}&1.3\tablenotemark{c}
 &2.2\tablenotemark{c}&\nodata&II\\
J1810+1744&1.7&4.0\tablenotemark{c}&0.9\tablenotemark{c}&1.8\tablenotemark{c}&
 \nodata&II\\
J1125$-$5825&3.1&8.1\tablenotemark{c}&4.4\tablenotemark{c}&
 1.4\tablenotemark{c}&\nodata&I\\
J1446$-$4701&2.2&3.7\tablenotemark{c}&1.5\tablenotemark{c}&1.3\tablenotemark{c}&
 \nodata&I\\
J2215+5135&2.6&5.2\tablenotemark{c}&2.5\tablenotemark{c}&1.3\tablenotemark{c}&
 \nodata&I\\
J2241$-$5236&2.2&3.3\tablenotemark{c}&1.4\tablenotemark{c}&1.2\tablenotemark{c}&
 \nodata&I\\
J1658$-$5324&2.4&3.0\tablenotemark{c}&1.7\tablenotemark{c}&1.1\tablenotemark{c}&
 \nodata&I\\
J0034$-$0534&1.9&1.7&0.75&1.0&\nodata&II\\
J1124$-$3653&2.4&1.6\tablenotemark{c}&1.2\tablenotemark{c}&0.78\tablenotemark{c}&
 \nodata&III\\
J0614$-$3329&3.2&2.2\tablenotemark{c}&2.4\tablenotemark{c}&0.71\tablenotemark{c}&
 \nodata&III\\
J2043+1711&2.4&1.3&1.0&0.70&\nodata&I\\
J0102+4839&3.0&1.8\tablenotemark{c}&1.9\tablenotemark{c}&0.67\tablenotemark{c}&
 \nodata&III\\
J2214+3000&3.1&1.9\tablenotemark{c}&2.2\tablenotemark{c}&0.66\tablenotemark{c}&
 \nodata&II/III\\
J1858$-$2216&2.4&1.1\tablenotemark{c}&1.0\tablenotemark{c}&0.66\tablenotemark{c}&
 \nodata&III\\
J0023+0923&3.1&1.5\tablenotemark{c}&1.8\tablenotemark{c}&0.60\tablenotemark{c}&
 \nodata&I\\
J2017+0603&2.9&1.3\tablenotemark{c}&1.6\tablenotemark{c}&0.59\tablenotemark{c}&
 \nodata&I\\
J1741+1351&3.8&2.2&3.3&0.58&\nodata&III\\
J0101$-$6422&2.6&1.0&1.1&0.58&\nodata&I\\
J0613$-$0200&3.1&1.2&1.7&0.53&\nodata&I\\
J1514$-$4946&3.6&1.6\tablenotemark{c}&2.6\tablenotemark{c}&0.52\tablenotemark{c}&
 \nodata&I\\
J0751+1807&3.5&0.72&1.7&0.36&\nodata&1\\
J0340+4130&3.3&0.65\tablenotemark{c}&1.4\tablenotemark{c}&0.36\tablenotemark{c}&
 \nodata& II/III\\
J2047+1053&4.3&1.1\tablenotemark{c}&3.0\tablenotemark{c}&0.35\tablenotemark{c}&
 \nodata&III\\
J1600$-$3053&3.6&0.73&1.8&0.35&\nodata&I\\
J1614$-$2230&3.2&0.38&1.0&0.29&\nodata&I\\
J1231$-$1411&3.7&0.51&1.6&0.29&\nodata&I\\
J2051$-$0827&4.5&0.54&2.4&0.24&\nodata&I\\
J1744$-$1134&4.1&0.41&1.7&0.23&\nodata&III\\
J1713+0747&4.6&0.34&2.0&0.19&\nodata&I\\
J2124$-$3358&4.9&0.37&2.4&0.18&BB&III\\
J0030+0451&4.9&0.35&2.3&0.18&BB+PL&I\\
J2302+4442&5.2&0.38\tablenotemark{c}&2.7\tablenotemark{c}&0.18\tablenotemark{c}&
 \nodata&I/II\\
J0437$-$4715&5.8&0.29&2.9&0.14&BB+PL&I\\
J0610$-$2100&3.9&0.08&0.69&0.11&\nodata&III\\
J1024$-$0719&5.2&0.05&0.92&0.06&\nodata&I
\enddata
\tablenotetext{a}{Power-law (PL) or blackbody (BB) models.}
\tablenotetext{b}{Alignment between the gamma-ray and radio pulse peaks.
Class I: the gamma-ray pulse lags the radio pulse; class II: the
gamma-ray and radio pulses are in phase; class III: the gamma-ray pulse
precedes the radio one. Both classes are listed if a pulsar cannot be
classified unambiguously.}
\tablenotetext{c}{Observed value --- there is no correction to \pdot\ for
the pulsar proper motion or differential Galactic rotation.}
\tablecomments{The pulsar parameters are all from the LAT second pulsar
catalog \citep{aaa+13}, except those of PSR B1821$-$24, which are from
\citet{jgk+13}.}
\end{deluxetable*}

In Table~\ref{table:fermimsp} we list all gamma-ray MSPs detected with \fer\
LAT in decreasing order of \blc. We classified them according to their phase
alignment between the gamma-ray and radio pulses, using a scheme similar to
\citet{vjh12}: sources with gamma-ray peak lagging, aligned with, and
preceding the radio peak are divided into classes I, II, and III,
respectively. Class II pulsars are required to have a radio peak aligned with
the gamma-ray peak to better than 1/10 of the spin period, and the two peaks
should have a similar profile. There are a small number of pulsars, e.g., PSR
J2214+3000, that cannot be unambiguously classified. These cases are noted in
the table. Finally, we also list in the table the X-ray spectral type of the
MSPs, although only a few of them are detected since MSPs are generally faint.
It is obvious that most high-\blc\ MSPs belong to class II and they exhibit
non-thermal X-ray emission with hard spectra.

\subsection{Modeling the Emission of High-\blc\ MSPs}
\citet{gjv+12} first noted that a group of MSPs show aligned radio, X-ray, and
gamma-ray pulse profiles, suggesting the same location for the emission. The
authors proposed that the radio emission could be generated in caustics in the
outer magnetosphere, same as the gamma-ray emission. As
Table~\ref{table:fermimsp} indicates, the phase alignment is typical among the
highest-\blc\ MSPs. We discuss below possible emission mechanisms in different
energy bands, with a focus on their connection with \blc. We qualitatively
compare our toy model with the spectral energy distributions (SEDs) of three
MSPs --- PSRs B1937+21, B1821$-$24, and J0218+4232, which have been detected
in both X-rays and gamma-rays. 

\subsubsection{Gamma-Ray Emission}
We briefly summarize the gamma-ray emission mechanism in the context of the
outer gap model \citep{chr86}, in which electrons and positrons are
accelerated up to a Lorentz factor of $\gamma\sim 10^7$ near the light
cylinder. For a more detailed calculation of the gamma-ray emission process,
we refer to \citet{wtc10,wtc11} and \citet{tct12}. In the outer gap, electrons
and positrons are accelerated by the electric field along the magnetic field
lines and emit GeV gamma-rays via the curvature radiation process. The
magnitude of the electric field is given by
\begin{equation}
E_\parallel\sim \frac{f_{\rm gap}B_{\rm lc}R_{\rm lc}}{R_c}~,
\end{equation}
where $R_c$ is the curvature radius and $f_{\rm gap}$ is the ratio between the
gap thickness and the light cylinder radius, typically $f_{\rm gap}\sim0.3$
for MSPs \citep{tct12}.

The Lorentz factor of the accelerated particles can be estimated by the
balance between the electric force and the back reaction force of the
curvature radiation,
\begin{eqnarray}
\gamma_p =& \left(\frac{3R_c^2}{2e}E_\parallel\right)^{1/4}
\sim 5\times 10^6\left(\frac{f_{\rm gap}}{0.3}\right)^{1/2}
\left(\frac{P}{1\,{\rm ms}}\right)^{1/2}\nonumber \\
&\times \left(\frac{B_{\rm lc}}{10^5\,{\rm G}}\right)^{1/4}
\left(\frac{R_{c}}{R_{\rm lc}}\right)^{1/4}~.
\end{eqnarray}
The energy of the curvature photons is of the order of GeV, 
\begin{eqnarray}
E_c=&\frac{3hc\gamma^3}{4\pi R_c}
\sim 0.8\left(\frac{f_{\rm gap}}{0.3}\right)^{1/4}
\left(\frac{P}{1\,{\rm ms}}\right)^{1/2}\nonumber \\
&\times \left(\frac{B_{\rm lc}}{10^5\,{\rm G}}\right)^{3/4}
\left(\frac{R_c}{R_{\rm lc}}\right)^{-1/4}\,{\rm GeV}
\end{eqnarray}
and the gamma-ray luminosity from the outer gap is typically
\begin{eqnarray}
L_{\gamma}\sim &f_{\rm gap}^3\dot E\sim 1.2\times 10^{32}
\left(\frac{f_{\rm gap}}{0.3}\right)^{3}
\left(\frac{P}{1\,{\rm ms}}\right)^{2}\nonumber \\
&\times \left(\frac{B_{\rm lc}}{10^5\,{\rm G}}\right)^{2}\,{\rm erg\,s^{-1}}~.
\end{eqnarray}

\subsubsection{X-Ray Emission}
In the standard pulsar theory, non-thermal X-ray emission is attributed to
synchrotron radiation from secondary pairs in the outer magnetosphere, which
are generated via the pair-creation process between gamma-rays and the
background X-rays \citep[e.g.,][]{tc07}. The secondary pairs have an initial
Lorentz factor of $\gamma_{\rm max}\sim (1\,{\rm GeV})/(2m_ec^2)\sim 10^3$,
then quickly lose their energy via synchrotron radiation and eventually leave
the light cylinder with $\gamma_{\rm min}\sim 1/\sin\theta_s\sim 10$, where
$\theta_s$ is the pitch angle. During this process, synchrotron cooling gives
non-thermal X-rays with $\Gamma_X\sim 1.5$ between $E_{\rm
min}\sim2\times10^{-2}\sin\theta_s (B_{lc}/10^5{\rm G})$\,keV to $E_{\rm
max}\sim5\sin\theta_s(B_{\rm lc}/10^5\,{\rm G})$\,keV. This is the general
case for young pulsars; they have $\Gamma_X$ observed in the range of 1.5--2
\citep[e.g.,][]{khc+01,hsg+02}.

For high-\blc\ MSPs, while non-thermal X-rays can be produced near the light
cylinder, the very hard photon indices ($\Gamma_X\sim1$) are difficult to
explain. A hard spectrum is expected below $E_{\rm min}$, but this is outside
the observation bands. \citet{hum05} proposed that the primary particles in
MSPs could maintain large momenta and undergo cyclotron resonant absorption of
radio emission to produce synchrotron X-rays with a very hard spectrum.
However, this process is not efficient between particles with $\gamma\sim10^7$
and radio waves of 0.1--1\,GHz, unless the $B$-field is of the order of
$10^7$--$10^8$\,G. Such a condition can only occur near the neutron star
surface.

One possible scenario to explain the non-thermal X-rays from the outer
magnetosphere is inverse-Compton (IC) scattering between the primary particles
and radio waves. As we discussed, the radio, X-ray, and gamma-ray emission
regions of high-\blc\ MSPs are likely co-located. Therefore, the radio waves
emitted in the outer magnetosphere may possibly irradiate the outer gap
region and be up-scattered by the ultra-relativistic particles. We can
estimate the energy density of the radio waves in the magnetosphere from the
flux density. Assuming a radio spectral index of $\alpha=2$, the typical
energy density of $\sim$100\,MHz radio waves is $U_{\rm ph}\sim10^4$--$10^
5$\,erg\,cm$^{-3}$ in high-\blc\ MSPs. The radiation power of IC scattering
from a single particle is then $P_{\rm IC}\sim4\sigma_Tc\gamma^2U_{\rm ph}/3$,
where $\sigma_T$ is the Thomson cross section. As a comparison, the power of
the curvature radiation is $P_{\rm cur}\sim2e^2c\gamma^4/3R_c^2$. Hence,
\begin{equation}
\frac{P_{\rm IC}}{P_{\rm cur}}\sim 0.1 \left(\frac{U_{\rm ph}}{10^4\,{\rm
erg\,cm^{-3}}}
\right)\left(\frac{\gamma}{10^7}\right)^{-2}\left(\frac{R_c}{10^7\,{\rm cm}}
\right)^2~.
\end{equation}
This suggests that the expected energy flux from the IC process is only
slightly smaller than that of the GeV emission and it could be observable
in high-\blc\ MSPs.

To model the IC spectrum, we consider the scattering between outgoing
particles in the outer gap and outwardly propagating radio waves. We assume
that the radio emission region lies just above the outer gap and approximate
the magnetic field lines by concentric circles. The collision angle between
the primary particles and radio waves can be crudely estimated by $\sin
\theta_0\sim\sqrt{2f_{\rm gap}}$. For each particle, the IC power per unit
energy per unit solid angle is given by
\begin{equation}
\frac{dP_{\rm IC}}{d\Omega}\sim {\mathcal D}^2(1-\beta\cos\theta_0)F_{\rm rad}
\frac{d\sigma'}{d\Omega'},
\end{equation} 
where $d\sigma'/d\Omega'$ is the differential Klein-Nishina cross section, 
$\beta=\sqrt{\gamma^2-1}/\gamma$, $\mathcal{D}=\gamma^{-1}(1-\beta\cos
\theta_1)^{-1}$, $0<\theta_1<1/\gamma$ is the angle between the particle
motion and the scattered photon direction, and $F_{\rm rad}$ is the radio
spectrum.

The IC spectrum depends sensitively on the radio spectrum. Although the latter
is not very clear at low radio frequency $<1$\,GHz, it is believed that a
spectral turnover should exist below 100\,MHz for MSPs \citep[see][]{kl01},
which is lower than that of young pulsars \citep[$\sim1$\,GHz; e.g.,
][]{kgk07}. For a turnover at 10--100\,MHz, the corresponding break in the IC
spectrum would be at $(1-10)\times (\gamma/10^7)^2$\,MeV. This is well above
the \cxo\ and \xmm\ energy bands, suggesting that the IC emission could
contributed to the observed X-rays. To compare with observations, we assume a
broken power-law spectrum in radio with a turnover at 10\,MHz, i.e.
\begin{equation}
F_{\rm rad}(\nu)=A\left \{ \begin{array}{@{\,}ll}
\left(\frac{\nu}{100\,{\rm MHz}}\right)^{\beta_1} & \mathrm{for}~ \nu\ge
10{\rm MHz} \\
\left(\frac{10\,{\rm MHz}}{100\,{\rm MHz}}\right)^{\beta_1}
\left(\frac{\nu}{10\,{\rm MHz}}\right)^{\beta_2} & \mathrm{for}~\nu< 10\,{\rm
MHz}.
\end{array} \right . \label{norm}
\end{equation}

The spectral index $\beta_1$ is inferred from the observed flux densities at
400\,MHz and 1.4\,GHz listed in the ATNF pulsar catalog \citep{mht+05}. The
index below 10\,MHz, $\beta_2$, is taken to be 0.5, which provides a good
match to the X-ray spectra (see below). The normalization $A$ is related to
the observed flux density at 400\,MHz ($F_{400}$) and the source distance $d$
by
\begin{equation}
A\sim F_{400}\left(\frac{d}{R_{\rm lc}}\right)^2
\left(\frac{100{\rm MHz}}{400{\rm MHz}}\right)^{\beta_1}~.
\end{equation}
As the exact geometry of the radio emission region is unknown, we allowed $A$
to vary by a factor of a few to fit the observed SEDs.

We also considered synchrotron X-rays contributed by the secondary pairs. The
particles accelerated towards the star would eventually reach the stellar
surface and heat up the polar cap region. \citet{tct12} estimated a surface
temperature of $\sim 1$\,MK and a luminosity of $L_r\sim5\times 10^{31}$\ergs\
for the thermal emission of PSR~B1937+21. These values are below our detection
limit of 1.5\,MK. The thermal X-rays may collide with the GeV gamma-rays to
create new pairs, which emit non-thermal X-rays via the synchrotron process.
The optical depth of the pair-creation process is
$\tau_{X\gamma}\sim L_r\sigma_{X\gamma}R_{\rm lc}/(4\pi R_{\rm lc}^2c kT_r)
\sim 0.02$, where $k_B$ is the Boltzmann constant and $\sigma_{X\gamma}\sim
\sigma_T/3$. The energy distribution of the secondary pairs follows
\begin{equation}
\frac{dN_e}{d\gamma_s}(\gamma_s)\sim \frac{m_ec^2}{\dot{E}_{\rm syn}}
 \int_{2\gamma_sm_ec^2}^{\infty}Q(E'_{\gamma}) dE'_{\gamma}~,
\label{sdist}
\end{equation}
where $\gamma_s$ is the Lorentz factor of the secondary pairs, $Q(E_{\gamma})
=F_{\rm cur}(1-e^{-\tau_{X\gamma}})/E_{\gamma}$, $F_{\rm cur}$ is the
curvature radiation power per unit energy, and $\dot{E}_{\rm syn}$ is the rate
of energy loss from synchrotron radiation. With $\dot{E}_{\rm
syn}=2e^4B^2\sin^2\theta_{s} \gamma_s^2/3m_e^2c^3$, Equation~(\ref{sdist})
describes a power-law with index $p=2$. The synchrotron spectrum from
secondary pairs is
\begin{equation}
F_{\rm syn}(E_{\gamma})=\frac{\sqrt{3}e^3B\sin\theta_s}{hm_ec^2}\int
\frac{dN_e}{d\gamma_s}F(x)d\gamma_s~,
\end{equation}
where $x=E_{\gamma}/E_{\rm syn}$, $E_{\rm syn}=3he\gamma_s^2B\sin\theta_s/4\pi
m_ec$, and $\sin \theta_s\sim \sqrt{2f_{\rm gap}}$.

\begin{figure*}[th]
\epsscale{0.7}
\plotone{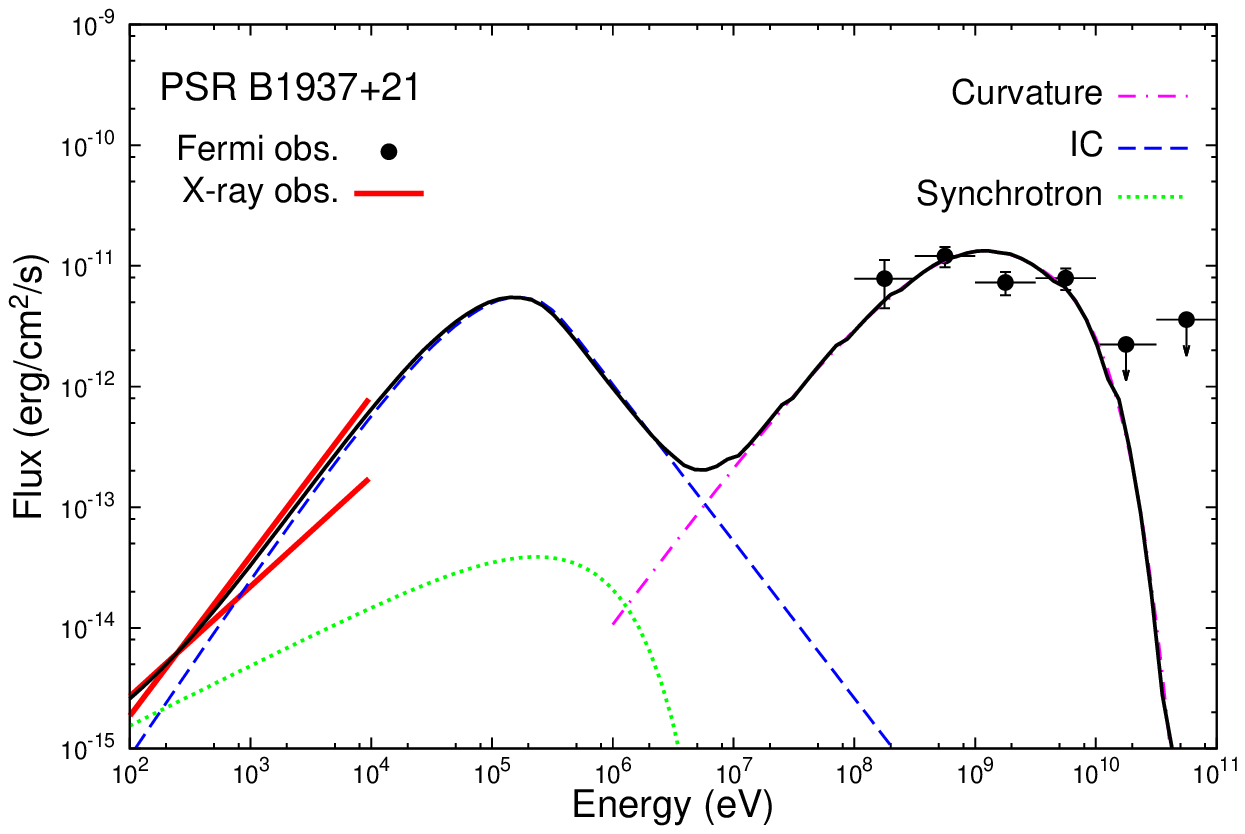}
\plotone{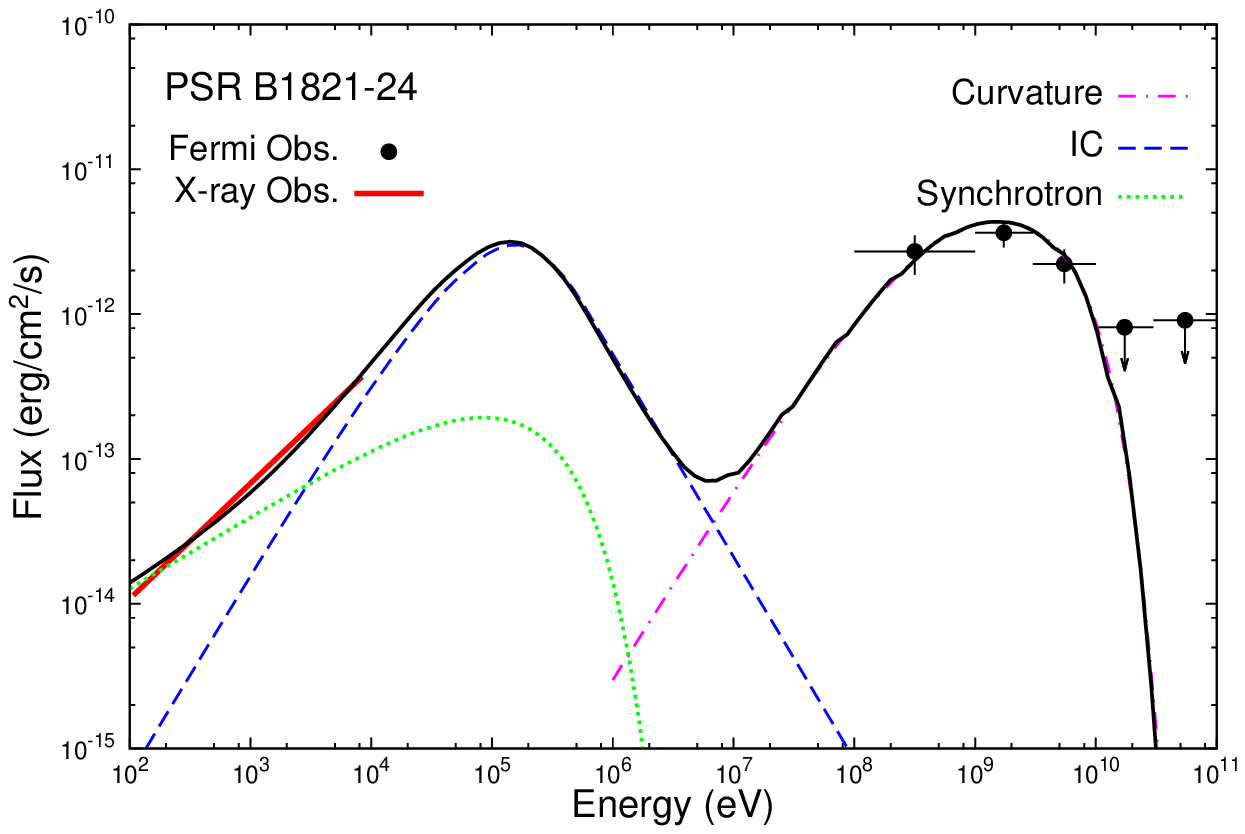}
\plotone{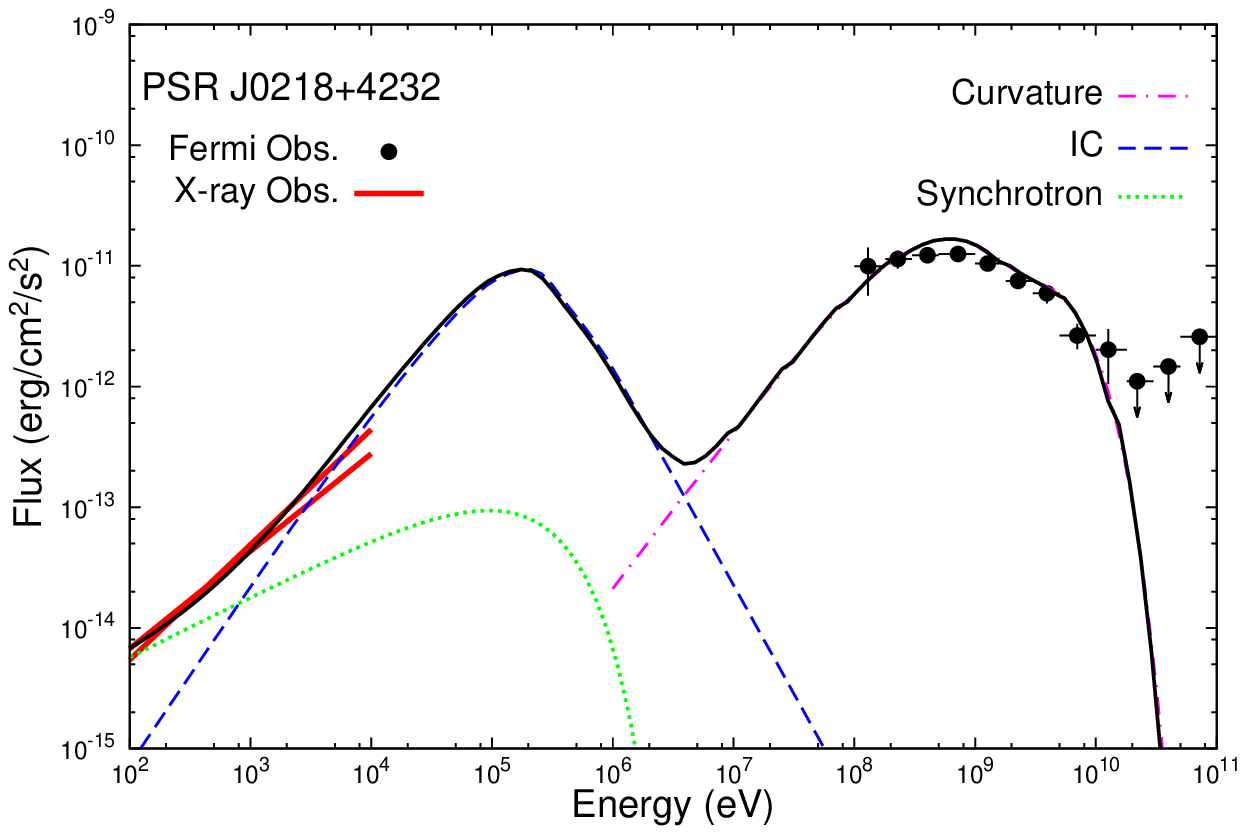}
\caption{Our emission model compared with the observed SEDs of PSRs
B1937+21, B1821$-$24, and J0218+4232 from X-rays to gamma-rays. The thick red
line and the dots show the X-ray and gamma-ray data, respectively.
The curvature radiation, IC emission, synchrotron emission, and total emission
of the model are indicated by the magenta dash-dot lines, blue dashed lines,
green dotted lines, and black lines, respectively. (A color version of this
figure is available in the online journal.)
\label{fig:sed}}
\end{figure*}

In Figure~\ref{fig:sed} we compare our simple model with the observed SEDs
of PSRs B1937+21, B1821$-$24, and J0218+4232. It shows that our model provides
reasonable fits to the data. The peak flux of the IC and curvature radiation
are similar and the IC emission dominates over the synchrotron radiation in
the X-ray band above a few keV. Since the IC spectrum is sensitive to the
radio spectral index, low-frequency radio measurements below 100\,MHz in
future can offer essential inputs to refine the modeling. For the assumed
spectral turnover at 10\,MHz, the IC emission peaks at $\sim$100\,keV, which
could be detectable with hard X-ray telescopes such as \emph{NuSTAR} or
\emph{ASTRO-H}.

\subsubsection{Radio Emission}
The radio emission process in pulsars is not clearly understood, nonetheless,
it has been suggested that plasma instability could play an important role in
the generation of the coherent radio emission \citep[e.g.,][]{uso87}. The
characteristic timescale for the instabilities (e.g., two-stream instability)
to develop is related to the inverse of the plasma frequency $\omega_p=\sqrt{
4\pi e^2n_e/m_e}$. The electron and positron number density, $n_e$, is
proportional to the Goldreich-Julian charge density ($n_{\rm GJ}$) such that
$n_e\propto n_{\rm GJ}\sim B_{\rm lc}/(2\pi Pc)$. Therefore, high-\blc\ MSPs
tend to have a higher plasma frequency, and hence a shorter instability
timescale, in the outer magnetosphere.

Above the outer gap accelerator, the outgoing and incoming gamma-rays produce
electron-positron pairs moving towards the light cylinder and the star,
respectively. These flows could develop two-stream instability in the outer
magnetosphere. Since there are more pairs created near the inner boundary than
near the outer boundary of the outer gap \citep{tsh04}, the outflow is
stronger than the inflow at the light cylinder. The Lorentz factor of the two
flows are of the order of $\gamma_s\sim 10$, for which the synchrotron cooling
timescale is comparable to light-crossing time of the magnetosphere. The
instability development timescale can be estimated by
\begin{equation}
\tau_i\sim \left(\frac{n_o}{n_i}\right)^{1/3}\gamma_s^{3/2}\omega_{p,o}^{-1}~,
\end{equation}
where $n_o$ and $n_i$ are the number densities of the outgoing and incoming
flows, respectively, and $\omega_{p,o}$ is the plasma frequency of the
outgoing flow \citep{uso87}. With a typical multiplicity of the order of 
$\kappa\sim 10^3$, we may assume $n_o\sim \kappa n_{\rm GJ}$ and $n_i\sim 
n_{\rm GJ}$ in the outer magnetosphere. Hence, the timescale becomes
\begin{equation}
\tau_{i}\sim 1.4\times 10^{-8}\left(\frac{\kappa}{10^{3}}\right)^{-1/6}
\left(\frac{P}{1{\,\rm ms}}\right)^{1/2}
\left(\frac{B_{\rm lc}}{10^5{\,\rm G}}\right)^{-1/2}{\,\rm s}~,
\end{equation}
which is much shorter than the light-crossing time of $\tau_c\sim R_{\rm
lc}/c\sim 1.6\times 10^{-4}(P/1{\,\rm ms})$\,s. As a result, the instability
could develop before the outgoing particles escape the magnetosphere. We
speculate that the two-stream instability may generate non-homogeneous and
separated plasma clouds, and the scattering or emission process of the plasma
cloud eventually produce the observed radio emission in phase with the
gamma-ray pulses.

Finally, we note that some high-\blc\ MSPs, e.g., PSRs B1821$-$24 and
J1810+1744, show complex radio profiles with additional components that offset
from the gamma-ray peaks. These could be contributed by emission from the
polar cap region as in the conventional theory of pulsar radio emission.

\section{CONCLUSIONS} \label{s5}
We have analyzed X-ray and gamma-ray observations of \psr\ taken with \cxo,
\xmm\ and \fer\ LAT. We obtained much improved spectral and timing
measurements than previous studies. Our results show that the pulsar X-ray
emission is $\sim$100\% pulsed and has a purely non-thermal spectrum that can
be described by a hard power-law of photon index $\Gamma_X=0.9\pm 0.1$. The
X-ray pulse profile consists of two sharp peaks $\sim$180\arcdeg\ apart. They
generally align with the radio peaks and the phase offsets are less than 7\%
of the spin period. In gamma-rays, the 5.5\,yr of \fer\ survey data provide a
good quality pulse profile in 0.1--100\,GeV with a significance of over
7$\sigma$. We performed a binned likelihood analysis on the pulsed emission
and found that a simple power-law model with $\Gamma_\gamma=2.38\pm0.07$ gives
a TS value of 112, corresponding to over 10$\sigma$ significance. Adding an
exponential cutoff to the power-law model slightly improves the fit, but the
change is not statistically significant.

A comparison of \psr\ with other MSPs indicates that sources with a strong
magnetic field at the light cylinder tend to show a hard, non-thermal X-ray
spectrum and good alignment of pulse profiles in different energy bands. The
latter suggests that the radio, X-ray, and gamma-ray emission could originate
from the same region in the outer magnetosphere. We speculate that radio
emission could be generated in the outer gap region when \blc\ is large, since
this could give rise to short instability time scales. We investigate a simple
model in which the non-thermal X-rays are contributed by IC scattering between
radio waves and primary particles in the outer magnetosphere and by
synchrotron radiation from secondary particles. We showed that this toy model
is capable to qualitatively reproduce the observed SEDs of the highest-\blc\
MSPs. Future observations at low radio frequencies and in hard X-rays can help
refine the modeling.

\acknowledgements
We thank the referee for careful reading and useful suggestions and thank
Vicky Kaspi and Anne Archibald for useful discussions. JT, GCKL, and KSC are
supported by a GRF grant of Hong Kong Government under HKU7009/11P.

{\it Facilities:}
\facility{CXO (ACIS)}, \facility{XMM (EPIC)}, \facility{Fermi (LAT)}

\end{document}